\documentclass[12pt,letterpaper]{article}

\usepackage[left=1in,right=1in,top=1in,bottom=1in]{geometry}
\usepackage{setspace}%%ensure a single instance only in the preamble

\usepackage{amsmath,amssymb,amsthm,mathtools,bm}

\usepackage{graphicx}
\usepackage{booktabs}
\usepackage{threeparttable}
\usepackage{longtable}
\usepackage{array}
\usepackage{multirow}
\usepackage{caption}
\usepackage{subcaption}
\usepackage{float}
\usepackage{placeins}
\usepackage{makecell}

\usepackage{enumitem}
\usepackage{xcolor}
\usepackage{appendix}
\usepackage{pdflscape}
\usepackage{csquotes}

\usepackage{chngcntr}
\counterwithin{figure}{section}
\counterwithin{table}{section}

\usepackage[authoryear]{natbib}

\usepackage[colorlinks=true,linkcolor=blue,citecolor=blue,urlcolor=blue]{hyperref}
\usepackage[nameinlink,capitalize]{cleveref}

\theoremstyle{definition}

\title{A Three-Variable Benchmark for\\Post-GFC Covered Interest Parity Deviations}
\author{Useong Shin\thanks{
		Sogang Business School, Sogang University (Seoul, Korea).\\
		ORCID: \href{https://orcid.org/0009-0003-0197-9003}{0009-0003-0197-9003}\\
		Email: \texttt{useong@sogang.ac.kr}
}}
\date{\today}

\begin{document}
	
	\maketitle
	\thispagestyle{empty}
	
	\begin{flushleft}
	\end{flushleft}
	
	% \noindent\textbf{Acknowledgments:}
	% leave blank
	
	\begin{abstract}
		This paper proposes a public daily-frequency benchmark for post-GFC government-bond CIP deviations. Although CIP deviations are observed daily, the literature lacks a canonical benchmark for daily regressions comparable to standard factor models in asset pricing. Using G10 plus KRW currency--tenor panels, I show that three lagged public state variables---NFCI, the nominal broad U.S. dollar index, and the Treasury 10-year minus 2-year slope---deliver strong in-sample and leave-one-year-out performance. Cointegration, quarter-end, and aggregation-difference diagnostics suggest that the benchmark captures a persistent background component rather than short-maturity quarter-end spikes or spurious level correlation.
	\end{abstract}
	
	\pagenumbering{arabic}
	
	\newpage
	% =================================================
	% Section: Introduction
	% =================================================
	\section{Introduction}
	\label{sec:intro}
	
	Covered interest parity (CIP) is a central no-arbitrage relation in
	international finance. In frictionless markets, spot exchange rates, forward
	exchange rates, and interest-rate differentials should rule out covered
	currency arbitrage. Post-Global Financial Crisis data show that this relation
	does not fully close. Cross-currency bases and government-bond CIP deviations
	are persistent, and are linked to intermediary balance-sheet capacity, dollar
	funding pressure, regulation, hedging demand, and broader macrofinancial
	conditions \citep{BPN08,BabPac09a,BabPac09b,BMMS16,DTV18,ADKS19,COZ21,RSS22}.
	
	This paper is motivated by a practical benchmarking gap. Daily CIP deviation
	panels are now available, but the empirical CIP literature has no standard
	daily-frequency public-data benchmark. This contrasts with empirical asset
	pricing, where standard factor models provide a common hurdle for new factors.
	Daily CIP regressions lack an analogous reference point. New funding,
	liquidity, regulatory, or intermediary-capital variables are therefore often
	judged against ad hoc specifications rather than against a transparent common
	benchmark.
	
	This gap reflects the nature of the existing determinants literature. Many
	important CIP variables are balance-sheet, liquidity, regulatory, or
	institutional objects. They are economically central, but often lower-frequency,
	difficult to construct daily, or not easily reproducible from public sources.
	The dependent variable is daily, while many natural explanatory variables are
	not. I ask whether a small set of public state variables can summarize daily
	government-bond CIP deviations well enough to serve as a reproducible benchmark.
	
	The benchmark uses three lagged macrofinancial variables: the Chicago Fed
	National Financial Conditions Index (NFCI), the nominal broad U.S. dollar
	index, and the Treasury 10-year minus 2-year yield slope. These variables
	summarize financial conditions, the global dollar cycle, and the U.S.
	yield-curve regime. I do not interpret them as new structural determinants of
	CIP deviations. They are public proxies for macrofinancial forces already
	emphasized in the literature. The contribution is to show that this compact
	specification provides a strong daily benchmark for the persistent component
	of post-GFC government-bond CIP deviations.
	
	The scope is intentionally narrow. I study government-bond CIP deviations at
	maturities of three months and longer. This differs from the ultra-short
	quarter-end effects emphasized by \citet{DTV18}, where one-week or one-month
	contracts can move sharply when they begin to cross quarter-end reporting
	dates. The data used here begin at the three-month tenor. The paper therefore
	does not aim to explain the sharpest quarter-end spikes. It studies the
	background component of government-bond CIP deviations visible across daily
	currency--tenor panels.
	
	The empirical analysis uses the government-bond CIP deviation data of
	\citet{DKSData25}. The sample runs from January 1, 2008 to June 30, 2025 and
	covers G10 currencies plus the Korean won. In each currency--tenor panel, I
	regress the CIP deviation on lagged NFCI, the lagged broad dollar index, and
	the lagged Treasury slope. The model uses no proprietary balance-sheet data,
	no currency-specific institutional variables, and no high-dimensional search
	over macrofinancial controls.
	
	The results are large relative to the simplicity of the specification. In
	separate currency--tenor regressions, the within-panel in-sample \(R^2\) is
	0.637. The leave-one-year-out (LOYO) pooled out-of-sample \(R^2\) is 0.499.
	These metrics remove currency--tenor mean differences and evaluate holdout
	observations against panel-specific estimation-sample means. The fit is
	therefore not driven mechanically by matching average levels across currencies
	or tenors.
	
	The result also survives more restrictive specifications. When tenors are
	stacked within each currency and the three state-variable slopes are shared
	across tenors, the in-sample \(R^2\) is 0.645 and the LOYO pooled \(R^2\) is
	0.607. Thus the fit is not simply a by-product of freely estimating every
	currency--tenor slope. A small public-data specification captures a substantial
	low-frequency component of post-GFC government-bond CIP deviations.
	
	The coefficient patterns are economically interpretable. NFCI is positive in
	most panels and is the most consistently significant regressor, consistent
	with a limits-to-arbitrage view. The Treasury slope also tends to enter
	positively, suggesting that the U.S. yield-curve regime is linked to the
	persistent component of government-bond CIP deviations. The broad dollar
	coefficient is more heterogeneous, consistent with a global state variable
	whose effect depends on currency-specific funding and hedging pressures.
	
	Several diagnostics clarify the scope of the benchmark. A short-tenor
	quarter-end dummy adds essentially no explanatory power, consistent with the
	three-month-and-longer maturity structure of the data. Cointegration and
	aggregation-difference diagnostics further suggest that the benchmark captures
	a persistent macrofinancial component rather than daily high-frequency noise or
	a purely spurious level relation. The specification should therefore be read
	as a benchmark for the background component of government-bond CIP deviations,
	not as a high-frequency arbitrage-execution model.
	
	The contribution is a benchmark contribution. First, the paper identifies the
	absence of a canonical daily public-data benchmark for CIP deviations. Second,
	it proposes a compact three-variable specification that is easy to reproduce.
	Third, it shows that the benchmark delivers strong in-sample and
	out-of-sample performance in government-bond CIP panels. Fourth, it provides a
	simple empirical hurdle: richer explanations of background government-bond CIP
	deviations should demonstrate incremental power relative to this benchmark.
	
	The remainder of the paper proceeds as follows. \Cref{sec:lit} reviews the
	related literature. \Cref{sec:data} describes the data and methodology.
	\Cref{sec:results} presents the baseline results. \Cref{sec:robust} reports
	the robustness diagnostics. \Cref{sec:discuss} discusses heterogeneity, regime
	dependence, and benchmark interpretation. \Cref{sec:conclusion} concludes.
	
	% =================================================
	% Section: Literature Review
	% =================================================
	\section{Literature Review}
	\label{sec:lit}
	
	This paper relates to three strands of literature: classical covered interest
	parity, post-GFC CIP deviations and intermediary constraints, and
	macrofinancial determinants of cross-currency bases. The common lesson is that
	CIP is a clean no-arbitrage relation in theory, but its enforcement in actual
	markets uses trading, funding, and balance-sheet capacity. Post-GFC CIP
	deviations are therefore not simple pricing errors. They are persistent price
	variables that reflect frictions in international financial markets.
	
	The paper builds on this view but has a different purpose. Existing studies
	have identified many relevant mechanisms: transaction costs, FX liquidity,
	dollar funding pressure, intermediary risk-taking capacity, regulation, central
	bank balance sheets, hedging demand, broad dollar conditions, and term-structure
	variables. This paper does not replace those explanations. It asks whether a
	small set of public daily state variables can provide a reproducible benchmark
	for government-bond CIP deviations.
	
	\subsection{Classical CIP and transaction costs}
	
	The early CIP literature asks whether covered interest arbitrage profits
	survive transaction costs in actual foreign exchange markets. \citet{FL75,FL77}
	show that many apparent arbitrage profits are greatly reduced once bid--ask
	spreads and other transaction costs are taken into account. \citet{Deardorff79}
	introduces one-way arbitrage, a weaker arbitrage condition that can still shape
	foreign exchange market equilibrium.
	
	Later work uses higher-quality and higher-frequency data. \citet{Taylor87},
	\citet{Clinton88}, and \citet{ARS08} show that CIP deviations are usually small
	in normal market conditions and tend to disappear quickly. These findings
	support the view that CIP is a strong pricing relation in liquid FX markets.
	
	The key implication is that CIP is not only an accounting identity. It is a
	tradable relation. The no-arbitrage condition is exact in frictionless theory,
	but actual arbitrage requires execution, funding, and balance-sheet capacity. A
	measured CIP deviation therefore need not be a risk-free profit. This paper
	carries this logic to the post-GFC government-bond setting, where deviations
	are more persistent and are plausibly linked to broader financial conditions.
	
	\subsection{Post-GFC deviations and intermediary constraints}
	
	The post-GFC literature changed the interpretation of CIP deviations. They are
	no longer viewed only as short-lived pricing errors. They can be persistent
	objects shaped by intermediary constraints and global dollar funding
	conditions. \citet{BPN08} and \citet{BabPac09a,BabPac09b} document large
	dislocations in FX swap and cross-currency swap markets during 2007--2008 and
	link them to dollar funding stress. \citet{MvP12} also emphasizes dollar
	shortages in the global banking system.
	
	Subsequent work shows that CIP deviations persist beyond crisis episodes.
	\citet{BMMS16} interpret cross-currency bases through balance-sheet costs,
	regulation, and dollar funding pressure. \citet{DTV18} document systematic
	post-GFC CIP deviations across major currencies and link them to intermediary
	balance-sheet constraints and quarter-end regulatory effects. \citet{ADKS19}
	connect CIP deviations to the dollar and bank leverage. More recent work,
	including \citet{RSS22} and \citet{BZN24}, also emphasizes limited
	balance-sheet capacity and imbalances in FX swap demand.
	
	This interpretation is closely related to limits-to-arbitrage and intermediary
	asset pricing. \citet{BrunPed09} show that market liquidity and funding
	liquidity can reinforce each other. \citet{GP11} show that margin constraints
	can generate law-of-one-price deviations. \citet{HK13} and \citet{AEM14}
	develop the view that intermediary capital and leverage matter for asset
	prices. The message is similar across these studies: even when a
	no-arbitrage relation is exact in theory, its enforcement depends on
	constrained intermediaries. If their funding and balance-sheet capacity are
	limited, deviations can persist.
	
	For this paper, the important implication is empirical. Many economically
	natural CIP variables are balance-sheet, regulatory, liquidity, or
	institutional objects. They are important, but they are often observed at lower
	frequencies, difficult to reconstruct daily, or not easily available from
	public sources. This creates a gap between daily CIP deviation panels and the
	data environment needed to test many structural mechanisms. A public daily
	benchmark is useful because it separates two questions: how much of
	government-bond CIP deviations is captured by easily observable macrofinancial
	state variables, and how much additional power is supplied by richer
	mechanism-specific variables.
	
	\subsection{Macrofinancial determinants and the benchmark role}
	
	The closest paper to this study is \citet{COZ21}. They study macrofinancial
	determinants of CIP deviations and show that FX market liquidity, intermediary
	risk-taking capacity, broad U.S. dollar strength, central bank balance sheets,
	and term premia are systematically related to cross-currency bases. This paper
	therefore does not claim that macrofinancial variables are new to the CIP
	literature. That connection is already well established.
	
	The difference is the role of the specification. Existing work identifies many
	variables related to CIP deviations, but these variables have not become a
	standard daily-frequency horse-race benchmark. In empirical asset pricing,
	standard factor models provide a common hurdle for new factors. Daily CIP work
	has no equally compact and reproducible public benchmark. A new explanatory
	variable can therefore look useful relative to an arbitrary straw-man model,
	even if it adds little beyond simple macrofinancial state variables.
	
	This paper is designed to provide such a hurdle. The benchmark uses only three
	lagged public variables: the Chicago Fed National Financial Conditions Index
	(NFCI), the nominal broad U.S. dollar index, and the Treasury 10-year minus
	2-year yield slope. These variables summarize financial conditions, the global
	dollar cycle, and the U.S. yield-curve regime. Each is motivated by prior work,
	but none is treated as a newly discovered structural determinant. The
	contribution is to combine them into a compact daily benchmark and to evaluate
	its in-sample and leave-one-year-out out-of-sample performance.
	
	\subsection{Government-bond CIP deviations}
	
	This paper focuses on government-bond CIP deviations. \citet{DS16} use
	local-currency government bonds and currency-hedged returns to study sovereign
	risk and frictions in international capital markets. \citet{DIS18} link the
	special role of U.S. Treasuries to the hedging demand of global investors.
	\citet{DS22} review the literature on CIP deviations, the dollar, and
	international capital-market frictions. The government-bond CIP deviation data
	used in this paper belong to this line of work \citep{DKS25,DKSData25}.
	
	Government-bond CIP deviations are related to money-market bases, but they also
	reflect sovereign bond markets, safe-asset demand, hedging demand, and the term
	structure of interest rates. This motivates the benchmark variables. NFCI
	captures broad financial conditions. The broad dollar index captures the global
	dollar regime. The Treasury slope captures the U.S. yield-curve environment.
	The resulting specification is not a short-term funding-spread model. It is a
	public macrofinancial state-variable benchmark for the persistent component of
	government-bond CIP deviations.
	
	The interpretation is empirical rather than structural. Significant
	coefficients should not be read as direct evidence for one specific mechanism.
	Instead, the results show that post-GFC government-bond CIP deviations are
	strongly aligned with a small set of public macrofinancial state variables.
	That makes the benchmark useful as a transparent hurdle for richer daily
	models of CIP deviations.
	
	% =================================================
	% Section: Data and Methodology
	% =================================================
	\section{Data and Methodology}
	\label{sec:data}
	
	This paper studies post-GFC government-bond covered interest parity (CIP)
	deviations. The dependent variable comes from the government-bond CIP deviation
	data of \citet{DKSData25}. For country \(i\), maturity \(n\), and date \(t\),
	the government-bond CIP deviation is
	\begin{equation}
		x^{Govt}_{i,n,t}
		=
		y^{Govt}_{i,n,t}
		-
		\rho_{i,n,t}
		-
		y^{Govt}_{USD,n,t}.
		\label{eq:cip_definition}
	\end{equation}
	Here \(y^{Govt}_{i,n,t}\) is the local-currency government-bond yield,
	\(\rho_{i,n,t}\) is the market-implied forward premium for hedging the
	currency-\(i\) cash flow into U.S. dollars, and \(y^{Govt}_{USD,n,t}\) is the
	matched-maturity U.S. Treasury yield. Thus, \(x^{Govt}_{i,n,t}\) is the
	difference between the synthetic dollar yield on a hedged foreign government
	bond and the direct dollar yield on a U.S. Treasury. I use the variable
	\texttt{cip\_govt}, reported in basis points.
	
	The sample runs from January 1, 2008 to June 30, 2025. It covers
	currency--tenor panels for G10 currencies plus the Korean won. G10 currencies
	form the core sample in the post-GFC CIP literature. The Korean won is added as
	a major open-economy currency from an advanced Asian financial market. I denote
	the government-bond CIP deviation for currency \(c\), tenor \(\tau\), and date
	\(t\) by \(CIP_{c,\tau,t}\).
	
	The benchmark uses three public macrofinancial state variables. The Chicago Fed
	National Financial Conditions Index (NFCI) summarizes broad financial
	conditions and intermediary constraints \citep{FRED_NFCI}. The nominal broad
	U.S. dollar index captures the global dollar cycle \citep{DTWEXBGS}. The
	Treasury 10-year minus 2-year yield slope summarizes the U.S. yield-curve
	regime \citep{FRED_DGS10,FRED_DGS2}.
	
	The variables are aligned with the information set available at date \(t\).
	NFCI is weekly. I use the previous weekly print and carry it forward to the
	daily CIP calendar. The broad dollar index is daily, but it can be missing on
	some CIP trading days. I carry forward the most recent available observation
	and then use a one-observation lag. The Treasury slope is constructed as
	\[
	Slope_{t-1}=DGS10_{t-1}-DGS2_{t-1},
	\]
	and is also lagged by one observation. Thus, \(NFCI_{t-1}\) denotes the latest
	lagged weekly NFCI print available at date \(t\), while \(Dollar_{t-1}\) and
	\(Slope_{t-1}\) denote the lagged broad dollar index and Treasury slope.
	
	I interpret these variables as compact state variables, not as separately
	identified structural channels. NFCI summarizes broad financial conditions
	rather than one specific friction. The Treasury slope may reflect term premia,
	monetary-policy expectations, business-cycle conditions, and long-bond supply
	and demand. The broad dollar index captures the global dollar environment. The
	goal is to build a low-dimensional public benchmark, not to replace richer
	mechanism-specific proxies.
	
	The main benchmark regression is
	\begin{equation}
		CIP_{c,\tau,t}
		=
		\alpha_{c,\tau}
		+
		\beta^{N}_{c,\tau} NFCI_{t-1}
		+
		\beta^{D}_{c,\tau} Dollar_{t-1}
		+
		\beta^{S}_{c,\tau} Slope_{t-1}
		+
		\varepsilon_{c,\tau,t}.
		\label{eq:baseline_ct}
	\end{equation}
	I estimate this equation separately for each currency--tenor panel. The
	specification is intentionally small: it uses three public state variables and
	panel-specific intercepts.
	
	In-sample fit is measured by a within-panel \(R^2\). For each
	currency--tenor panel, I compute the total sum of squares around that panel's
	own mean. I then sum residual and total sums of squares across panels. This
	prevents the statistic from mechanically rewarding the model for matching
	average level differences across currencies or tenors.
	
	Out-of-sample performance is evaluated by leave-one-year-out (LOYO) tests. I
	exclude one calendar year at a time, estimate the model on the remaining years,
	and predict the excluded year. The LOYO \(R^2\) compares holdout observations
	with the training-sample mean within the same currency--tenor panel. This
	benchmark already allows each panel to have its own average level.
	
	I also estimate two stacked specifications. The first is a currency-level
	common-slope specification. For each currency \(c\), I stack all tenors and
	include tenor fixed effects:
	\begin{equation}
		CIP_{c,\tau,t}
		=
		\alpha_c
		+
		\lambda_{\tau,c}
		+
		\beta^{N}_{c} NFCI_{t-1}
		+
		\beta^{D}_{c} Dollar_{t-1}
		+
		\beta^{S}_{c} Slope_{t-1}
		+
		u_{c,\tau,t}.
		\label{eq:currency_common}
	\end{equation}
	This specification allows tenor-level differences within each currency, but
	restricts the three state-variable slopes to be common across tenors.
	
	The second is a pooled common-slope specification. I stack all currencies and
	tenors and include currency and tenor fixed effects:
	\begin{equation}
		CIP_{c,\tau,t}
		=
		\alpha
		+
		\mu_c
		+
		\lambda_{\tau}
		+
		\beta^{N} NFCI_{t-1}
		+
		\beta^{D} Dollar_{t-1}
		+
		\beta^{S} Slope_{t-1}
		+
		v_{c,\tau,t}.
		\label{eq:pooled_common}
	\end{equation}
	Together, the stacked specifications test whether the benchmark's performance
	depends on freely estimated currency--tenor slopes.
	
	For inference, I use Newey--West HAC standard errors. The daily level
	regressions use a 21-trading-day lag, roughly one trading month. In stacked
	regressions, many currencies and tenors are observed on the same date. I
	therefore aggregate scores by date before computing the HAC covariance matrix,
	which reduces the risk that cross-sectional stacking understates standard
	errors.
	
	I use two residual-based Engle--Granger diagnostics. The first tests the
	relation between actual CIP deviations and baseline fitted values. The second
	tests the relation between actual CIP deviations and the three baseline
	regressors directly. These tests are not causal evidence. They are diagnostic
	checks against purely spurious level correlation.
	
	Finally, I run an aggregation-difference analysis. For each currency--tenor
	panel, I sort observations by date and form non-overlapping \(N\)-trading-day
	blocks. I average the CIP deviation and the three regressors within each block,
	then take differences across adjacent blocks:
	\begin{equation}
		\Delta_N CIP_{c,\tau,b}
		=
		a_{c,\tau}
		+
		b^{N}_{c,\tau} \Delta_N NFCI_b
		+
		b^{D}_{c,\tau} \Delta_N Dollar_b
		+
		b^{S}_{c,\tau} \Delta_N Slope_b
		+
		e_{c,\tau,b},
		\label{eq:aggdiff}
	\end{equation}
	where \(b\) indexes non-overlapping aggregation blocks. This design asks
	whether the benchmark explains only persistent levels or also lower-frequency
	changes. Because the blocks do not overlap, it reduces mechanical serial
	correlation from overlapping differences. HAC lags for the differenced
	regressions use a conservative multiple of an automatic lag rule.
	
	The empirical design therefore has three steps. I first estimate a compact
	level benchmark using public state variables. I then evaluate out-of-sample
	performance and test whether slopes transport across tenors. Finally, I use
	cointegration and aggregation-difference diagnostics to assess whether the
	relation reflects stable low-frequency alignment rather than a spurious level
	fit.

	% =================================================
	% Section: Empirical Results
	% =================================================
	\section{Empirical Results}
	\label{sec:results}
	
	This section evaluates the three-variable benchmark. The model is not a
	structural explanation of CIP deviations. Its purpose is more limited: to test
	whether lagged NFCI, the lagged broad dollar index, and the lagged Treasury
	10-year minus 2-year slope summarize a large component of post-GFC
	government-bond CIP deviations.
	
	The results are strong for such a small specification. The benchmark delivers
	high in-sample fit, survives leave-one-year-out (LOYO) out-of-sample tests, and
	has economically interpretable coefficient patterns.
	
	\subsection{Overall fit}
	
	\Cref{tab:main_fit_summary} reports the main fit statistics. In separate
	currency--tenor regressions, the within-panel in-sample \(R^2\) is 0.637. This
	is a conservative statistic. It removes average level differences within each
	currency--tenor panel, so the fit is not driven mechanically by matching
	currency or tenor means.
	
	Out-of-sample performance is also strong. The separate currency--tenor
	specification has a LOYO pooled \(R^2\) of 0.499. The simple average of
	holdout-year \(R^2\) values is 0.347. Thus, coefficients estimated without a
	given calendar year still predict that excluded year better than the
	training-sample panel mean.
	
	The fit does not rely only on freely estimated currency--tenor slopes. In the
	currency-specific common-slope specification, all tenors are stacked within each
	currency and tenor fixed effects are included. This specification has an
	in-sample \(R^2\) of 0.645 and a LOYO pooled \(R^2\) of 0.607. The pooled
	common-slope specification, with currency and tenor fixed effects, has an
	in-sample \(R^2\) of 0.658.
	
	\FloatBarrier
	\begin{table}[H]
		\centering
		\onehalfspacing
		\footnotesize
		\caption{Baseline fit and out-of-sample performance}
		\label{tab:main_fit_summary}
		\begin{threeparttable}
			\begin{tabular}{lccc}
				\toprule
				Specification & In-sample \(R^2\) & LOYO pooled \(R^2\) & LOYO mean-year \(R^2\) \\
				\midrule
				Separate currency--tenor regressions & 0.637 & 0.499 & 0.347 \\
				Currency-specific common-slope regressions & 0.645 & 0.607 & 0.509 \\
				Pooled common-slope regression & 0.658 & -- & -- \\
				\bottomrule
			\end{tabular}
			\begin{tablenotes}
				\footnotesize
				\item Notes: The baseline specification uses lagged NFCI, the lagged nominal broad dollar index, and the lagged Treasury 10-year minus 2-year yield slope. Separate currency--tenor regressions estimate one regression for each currency--tenor panel. Currency-specific common-slope regressions stack all tenors within each currency and include tenor fixed effects. The pooled common-slope regression stacks all currencies and tenors and includes currency and tenor fixed effects. LOYO denotes leave-one-year-out evaluation. In-sample \(R^2\) uses a within-panel denominator. LOYO \(R^2\) uses the training-sample mean as the benchmark. The pooled common-slope specification is reported as an in-sample diagnostic only; LOYO evaluation is not reported for this fully pooled specification.
			\end{tablenotes}
		\end{threeparttable}
	\end{table}
	\FloatBarrier
	
	The maturity profile is also informative. \Cref{tab:tenor_profile} summarizes
	the separate currency--tenor fits by tenor. The benchmark is strongest at
	intermediate maturities. Mean \(R^2\) rises from 0.391 at the 3-month tenor to
	0.615 at the 5-year tenor, and remains high at the 7-year tenor. Fit is weaker
	at the very short end and at the 20-year tenor. This pattern suggests that the
	benchmark is not just capturing short-horizon money-market noise. It is most
	closely aligned with the middle part of the government-bond CIP term structure.
	
	\FloatBarrier
	\begin{table}[H]
		\centering
		\onehalfspacing
		\footnotesize
		\caption{Tenor profile of separate currency--tenor fits}
		\label{tab:tenor_profile}
		\begin{threeparttable}
			\begin{tabular}{lrrrrc}
				\toprule
				Tenor & Currencies & Observations & Mean \(R^2\) & Median \(R^2\) & Range of \(R^2\) \\
				\midrule
				0.25 & 11 & 49,205 & 0.391 & 0.379 & [0.160, 0.696] \\
				1    & 11 & 48,036 & 0.454 & 0.409 & [0.182, 0.887] \\
				2    & 11 & 48,468 & 0.522 & 0.498 & [0.219, 0.903] \\
				3    & 11 & 48,265 & 0.585 & 0.608 & [0.284, 0.871] \\
				5    & 11 & 48,404 & 0.615 & 0.631 & [0.254, 0.825] \\
				7    & 11 & 48,339 & 0.563 & 0.610 & [0.286, 0.782] \\
				10   & 11 & 48,395 & 0.465 & 0.421 & [0.156, 0.757] \\
				20   & 11 & 43,892 & 0.378 & 0.297 & [0.147, 0.682] \\
				30   & 10 & 36,809 & 0.421 & 0.505 & [0.149, 0.665] \\
				\bottomrule
			\end{tabular}
			\begin{tablenotes}
				\footnotesize
				\item Notes: The table summarizes full-sample separate currency--tenor regressions by tenor. For each tenor, \(R^2\) statistics are computed across available currencies. The observations column reports the total number of observations across currencies at each tenor. The 30-year tenor is available for 10 currencies.
			\end{tablenotes}
		\end{threeparttable}
	\end{table}
	\FloatBarrier
	
\begin{figure}[H]
		\centering
		\includegraphics[width=6.5in]{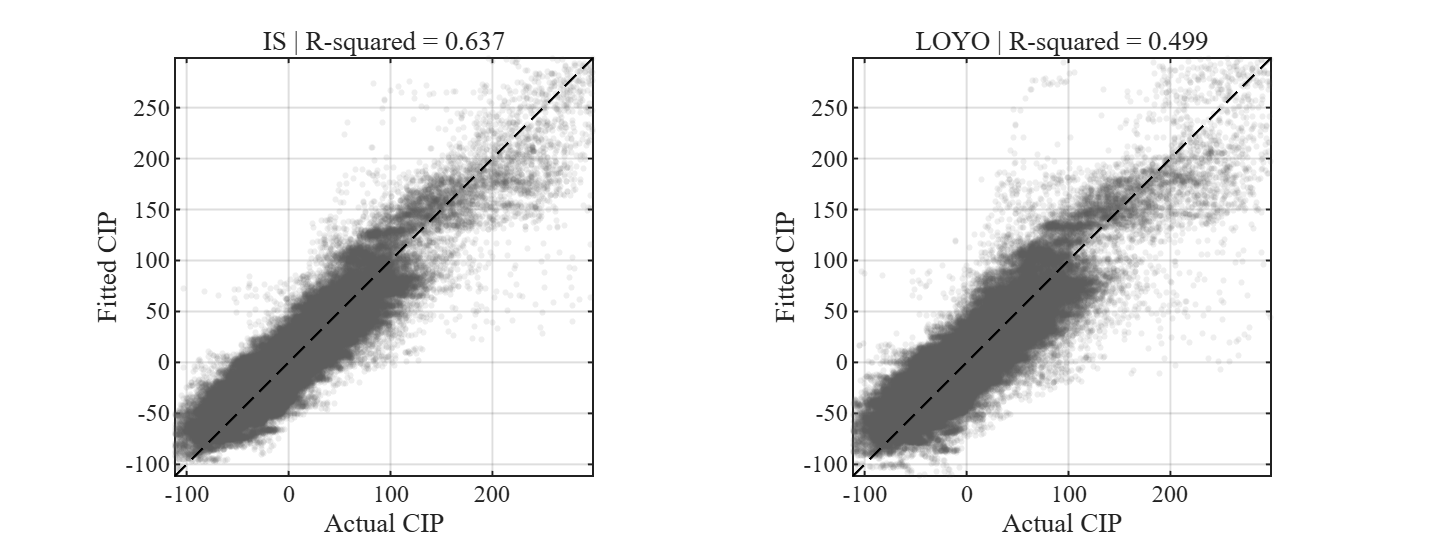}
		\caption{Actual CIP deviations and baseline fitted values. The out-of-sample fitted values are generated from the leave-one-year-out specification.}
		\label{fig:scatter}
	\end{figure}

	\Cref{fig:scatter} plots actual CIP deviations against fitted values. The
	figure shows a strong common component, but also substantial residual
	dispersion. This is the intended interpretation. The benchmark captures a large
	macrofinancial component of government-bond CIP deviations, but it is not a
	complete structural model.
	
	\FloatBarrier
	\FloatBarrier
	
	\subsection{Out-of-sample performance}
	
	\subsubsection{Leave-one-year-out validation}
	
	The first out-of-sample exercise is LOYO validation. This is not a fully
	chronological forecast, because the training sample for a holdout year can
	include later observations. Its purpose is different. It applies the same
	exclusion rule to every calendar year and tests whether the fit is concentrated
	in a few favorable episodes.
	
	\FloatBarrier
	\begin{figure}[H]
		\centering
		\includegraphics[width=6.5in]{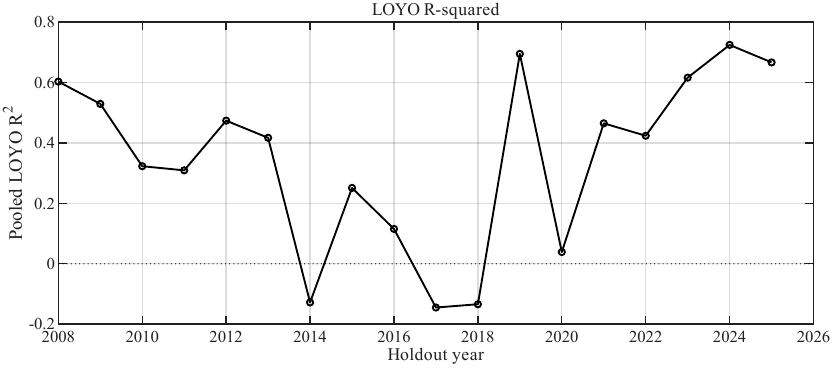}
		\caption{Leave-one-year-out out-of-sample performance}
		\label{fig:loyo}
	\end{figure}
	\FloatBarrier
	
	\Cref{fig:loyo} reports LOYO performance by holdout year. The separate
	currency--tenor specification delivers positive pooled \(R^2\) values in most
	years. The exceptions are 2014, 2017, and 2018. Thus, the benchmark does not
	fit every market environment mechanically. Even so, the full-sample LOYO pooled
	\(R^2\) remains high at 0.499.
	
	The currency-specific common-slope specification is more stable. It estimates
	common state-variable slopes within each currency and allows tenor fixed
	effects. Its LOYO pooled \(R^2\) is 0.607. This shows that substantial
	out-of-sample fit remains even when the model does not freely estimate every
	currency--tenor slope.
	
	\Cref{tab:currency_common_performance} reports the currency-specific
	common-slope results by currency. Most currencies show strong in-sample and
	out-of-sample performance. CAD, EUR, KRW, DKK, and SEK are especially strong.
	JPY is weaker. The key point is that the benchmark is not driven by one or two
	currencies.
	
	\FloatBarrier
	\begin{table}[H]
		\centering
		\onehalfspacing
		\footnotesize
		\caption{Currency-specific common-slope performance}
		\label{tab:currency_common_performance}
		\begin{threeparttable}
			\begin{tabular}{lrrr}
				\toprule
				Currency & Observations & In-sample \(R^2\) & LOYO pooled \(R^2\) \\
				\midrule
				AUD & 37,945 & 0.497 & 0.350 \\
				CAD & 39,842 & 0.799 & 0.776 \\
				CHF & 39,750 & 0.551 & 0.486 \\
				DKK & 38,987 & 0.595 & 0.546 \\
				EUR & 39,898 & 0.651 & 0.603 \\
				GBP & 39,563 & 0.491 & 0.429 \\
				JPY & 38,252 & 0.242 & 0.134 \\
				KRW & 35,400 & 0.746 & 0.746 \\
				NOK & 37,163 & 0.497 & 0.404 \\
				NZD & 34,550 & 0.581 & 0.529 \\
				SEK & 38,463 & 0.630 & 0.544 \\
				\bottomrule
			\end{tabular}
			\begin{tablenotes}
				\footnotesize
				\item Notes: For each currency, all tenors are stacked. The specification includes tenor fixed effects and common slopes on lagged NFCI, the lagged nominal broad dollar index, and the lagged Treasury 10-year minus 2-year yield slope. LOYO \(R^2\) uses the training-sample currency mean as the benchmark.
			\end{tablenotes}
		\end{threeparttable}
	\end{table}
	\FloatBarrier
	
	\subsubsection{Expanding-window forecasts}
	
	As a stricter time-ordered check, I also estimate expanding-window forecasts.
	The initial training window uses the first three sample years. The model is
	then re-estimated recursively using only past data, and the fitted coefficients
	are used to predict the next calendar year. This design avoids using future
	observations when evaluating a holdout year.
	
\begin{figure}[H]
		\centering
		\includegraphics[width=6.5in]{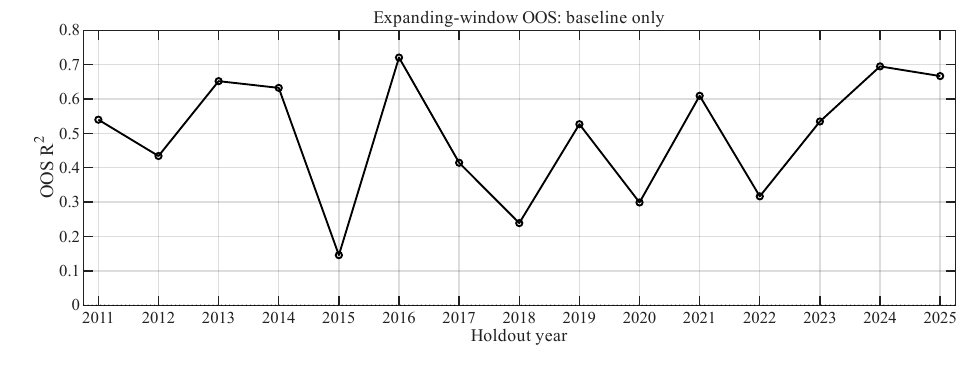}
		\caption{Expanding-window out-of-sample performance}
		\label{fig:expanding}
	\end{figure}

	\Cref{fig:expanding,tab:expanding_performance} report the results. The baseline
	remains strong under this chronological design. The expanding-window pooled
	\(R^2\) is 0.503 over 2011--2025, close to the LOYO pooled \(R^2\) of the
	separate currency--tenor specification. The simple average of yearly pooled
	\(R^2\) values is 0.495. All holdout years have positive pooled \(R^2\). The
	weakest years are 2015, 2018, 2020, and 2022, but even these remain positive on
	the pooled metric.
	
	\FloatBarrier
	\FloatBarrier
	
	\FloatBarrier
	\begin{table}[H]
		\centering
		\onehalfspacing
		\footnotesize
		\caption{Expanding-window out-of-sample performance by holdout year}
		\label{tab:expanding_performance}
		\begin{threeparttable}
			\begin{tabular}{lrrrr}
				\toprule
				Holdout year & Observations & Panels & Pooled \(R^2\) & Median panel \(R^2\) \\
				\midrule
				2011 & 22,906 & 90 & 0.540 & 0.465 \\
				2012 & 23,757 & 92 & 0.434 & 0.564 \\
				2013 & 23,806 & 92 & 0.652 & 0.395 \\
				2014 & 25,019 & 97 & 0.632 & 0.472 \\
				2015 & 24,541 & 97 & 0.146 & -0.894 \\
				2016 & 24,390 & 97 & 0.720 & 0.225 \\
				2017 & 24,251 & 97 & 0.414 & -0.119 \\
				2018 & 24,191 & 97 & 0.239 & 0.079 \\
				2019 & 24,092 & 97 & 0.527 & 0.309 \\
				2020 & 24,042 & 97 & 0.299 & 0.206 \\
				2021 & 23,685 & 97 & 0.609 & 0.438 \\
				2022 & 22,667 & 97 & 0.317 & 0.342 \\
				2023 & 23,181 & 97 & 0.534 & 0.424 \\
				2024 & 23,229 & 97 & 0.695 & 0.560 \\
				2025 & 11,463 & 98 & 0.667 & 0.456 \\
				\midrule
				All years & 345,220 & 1,439 & 0.503 & -- \\
				Mean year & -- & -- & 0.495 & 0.261 \\
				\bottomrule
			\end{tabular}
			\begin{tablenotes}
				\footnotesize
				\item Notes: The initial training window uses the first three sample years. The model is re-estimated recursively using only past data to predict each subsequent holdout year. Pooled \(R^2\) aggregates squared forecast errors across all test observations in the holdout year. Median panel \(R^2\) is the median across currency--tenor panels within the holdout year. The all-years pooled \(R^2\) aggregates all expanding-window test observations.
			\end{tablenotes}
		\end{threeparttable}
	\end{table}
	\FloatBarrier
	
	Panel-level performance is more heterogeneous. The median currency--tenor panel
	\(R^2\) is positive in 13 of the 15 expanding-window holdout years, with an
	average yearly median panel \(R^2\) of 0.261. This median statistic is useful
	because some currency--tenor panels produce very poor year-specific forecasts,
	which makes the simple mean of panel \(R^2\) values sensitive to outliers. The
	expanding-window evidence therefore supports the main result, while also
	showing that the benchmark is a compact macrofinancial summary rather than a
	complete panel-by-panel forecasting model.
	
	\subsection{Coefficient structure}
	
	\Cref{tab:coef_significance} summarizes coefficient signs and significance from
	98 separate currency--tenor regressions. NFCI is the most stable variable. Its
	coefficient is positive in 88.8\% of panels and significant at the 5\% level in
	86.7\%. Among significant NFCI coefficients, 80 are positive and only 5 are
	negative. This is consistent with the view that tighter financial conditions
	widen CIP deviations.
	
	The Treasury slope also has a stable sign pattern. Its coefficient is positive
	in 85.7\% of panels and significant at the 5\% level in half of the panels.
	Almost all significant slope coefficients are positive. This suggests that the
	U.S. yield-curve regime explains part of the persistent component not captured
	by NFCI and the broad dollar.
	
	The broad dollar coefficient is more heterogeneous. It is significant at the
	5\% level in 37.8\% of panels, but its sign varies across currencies and
	tenors. This is consistent with the broad dollar acting as a global state
	variable whose effect depends on currency-specific funding and hedging
	pressures.
	
	\FloatBarrier
	\begin{table}[H]
		\centering
		\onehalfspacing
		\footnotesize
		\caption{Coefficient signs and significance in separate currency--tenor regressions}
		\label{tab:coef_significance}
		\begin{threeparttable}
			\setlength{\tabcolsep}{4.5pt}
			\begin{tabular}{lrrrrrr}
				\toprule
				Variable 
				& \makecell{Mean\\coefficient} 
				& \makecell{Median\\coefficient} 
				& \makecell{Positive\\share} 
				& \makecell{5\% significant\\share} 
				& \makecell{Significant\\positive} 
				& \makecell{Significant\\negative} \\
				\midrule
				NFCI           & 22.954 & 18.808 & 0.888 & 0.867 & 80 & 5 \\
				Broad dollar   & -0.037 & -0.090 & 0.429 & 0.378 & 23 & 14 \\
				Treasury slope & 9.623  & 10.312 & 0.857 & 0.500 & 48 & 1 \\
				\bottomrule
			\end{tabular}
			\begin{tablenotes}
				\footnotesize
				\item Notes: The table summarizes coefficients from 98 separate currency--tenor regressions. The 5\% significant share is the fraction of panels with HAC \(p\)-values below 0.05. Significant positive and significant negative count coefficients that are positive or negative and significant at the 5\% level.
			\end{tablenotes}
		\end{threeparttable}
	\end{table}
	\FloatBarrier
	
	Overall, the results deliver three messages. First, three public state variables
	explain a large part of post-GFC government-bond CIP deviations. Second, the
	fit is not driven only by separate currency--tenor slopes; it remains strong in
	common-slope specifications. Third, the coefficient structure is economically
	sensible. NFCI and the Treasury slope show stable positive patterns, while the
	broad dollar captures heterogeneous exposure to the global dollar regime. The
	baseline is therefore a simple and reproducible public-data benchmark for
	future work on CIP deviations.
	
	\subsection{Economic magnitude}
	
	The coefficient patterns above show that the baseline variables are statistically
	stable, especially NFCI and the Treasury slope. I next translate the coefficients
	into basis-point magnitudes. For each state variable \(X\), I compute the effect
	of a one-standard-deviation move as
	\[
	\widehat{\beta}^{X}_{c,\tau}\times \sigma(X),
	\]
	and the effect of an interquartile-range move as
	\[
	\widehat{\beta}^{X}_{c,\tau}\times IQR(X).
	\]
	The table reports the median effect across the 98 separate currency--tenor
	regressions, with the cross-panel interquartile range of the effect in brackets.
	
	\FloatBarrier
	\begin{table}[H]
		\centering
		\onehalfspacing
		\footnotesize
		\caption{Economic magnitudes of baseline state variables}
		\label{tab:economic_magnitude}
		
		\resizebox{6.5in}{!}{%
			\begin{tabular}{lrrrrr}
				\toprule
				Variable
				& State SD
				& State IQR
				& Median coefficient
				& 1-SD effect
				& IQR effect \\
				\midrule
				NFCI
				& 0.605
				& 0.299
				& 18.808
				& 11.382 [5.403, 17.276]
				& 5.621 [2.669, 8.532] \\
				Broad dollar
				& 12.395
				& 22.846
				& -0.090
				& -1.122 [-7.301, 10.586]
				& -2.068 [-13.456, 19.511] \\
				Treasury slope
				& 0.968
				& 1.540
				& 10.312
				& 9.982 [4.351, 13.336]
				& 15.880 [6.921, 21.216] \\
				\bottomrule
			\end{tabular}%
		}
		
		\vspace{0.5em}
		
		\parbox{6.5in}{%
			\footnotesize
			\textit{Notes:} Effects are measured in basis points of government-bond CIP
			deviations. State-variable standard deviations and interquartile ranges are
			computed in the matched regression sample. The 1-SD and IQR effects multiply
			each panel-specific coefficient by the corresponding state-variable standard
			deviation or interquartile range. Brackets report the 25th and 75th percentiles
			across separate currency--tenor regressions.
		}
	\end{table}
	\FloatBarrier
	
	The magnitudes are economically meaningful. A one-standard-deviation increase in
	NFCI is associated with a median increase of about 11.4 basis points in
	government-bond CIP deviations. The Treasury slope has a similar
	one-standard-deviation effect of about 10.0 basis points. Its interquartile-range
	effect is even larger, about 15.9 basis points. Thus, the two most stable
	regressors in the coefficient-sign analysis are not only statistically reliable;
	they also have economically sizable effects.
	
	The broad dollar has a smaller median effect. A one-standard-deviation move in
	the broad dollar index is associated with a median effect of about -1.1 basis
	points. This small median should not be read as evidence that the dollar state
	variable is irrelevant. It reflects the heterogeneous sign pattern documented
	above. The interquartile range of the one-standard-deviation effect runs from
	about -7.3 to 10.6 basis points across panels. This is consistent with the
	interpretation that broad-dollar exposure differs across currencies and tenors,
	whereas NFCI and the Treasury slope capture more stable common components.
	
	% =================================================
	% Section: Robustness
	% =================================================
	\section{Robustness}
	\label{sec:robust}
	
	This section clarifies the scope of the benchmark through five robustness
	exercises. First, I test whether the results are affected by the quarter-end
	regulatory effects emphasized in the post-GFC CIP literature. Second, I compare
	the baseline with alternative VIX specifications to examine whether the benchmark
	is mainly capturing standard equity-market risk sentiment. Third, I use
	Engle--Granger diagnostics to assess whether the level fit is only a
	common-trend artifact. Fourth, I use non-overlapping aggregation differences to
	ask whether the relation also appears in lower-frequency changes. Fifth, I rotate
	the three state variables into principal components to test whether the benchmark
	is merely a proxy for one or two broad common factors.
	
	The results support a narrow interpretation. The benchmark is not a model of
	ultra-short quarter-end spikes, daily execution noise, or a single generic
	risk-sentiment factor. It is a compact public-data benchmark for the persistent
	macrofinancial component of government-bond CIP deviations.
	
	\subsection{Quarter-end dummy extension}
	
	A prominent result in the post-GFC CIP literature is that CIP deviations move
	around quarter ends. \citet{DTV18} show that the effect is especially sharp at
	very short maturities. One-month deviations rise when one-month contracts begin
	to cross quarter-end reporting dates, and one-week deviations rise when
	one-week contracts cross quarter ends.
	
	The present data are not designed to isolate that mechanism. The
	government-bond CIP deviation data of \citet{DKSData25} start at the
	three-month tenor. Three-month contracts are less suited for detecting the
	discrete quarter-end crossing effect, because they appear on quarter-end
	balance sheets regardless of the initiation date within the quarter.
	
	To check whether this matters for the benchmark, I augment the baseline with a
	short-tenor quarter-end dummy. The main version uses the last five observed
	trading days of each calendar quarter:
	\[
	QEndShort^{(5)}_{\tau,t}
	=
	\mathbf{1}\{\tau \leq 1\}
	\mathbf{1}\{t \in QEnd^{(5)}\},
	\]
	where \(QEnd^{(5)}\) denotes the last five observed trading days of each calendar
	quarter. The dummy is active only for tenors of one year or less. This window
	allows quarter-end balance-sheet pressure to be reflected gradually before the
	reporting date. I also check a narrower version that activates only on the final
	observed trading day of each quarter.
	
	The results are economically negligible. The five-day quarter-end dummy changes
	the in-sample \(R^2\) by less than 0.0001 and does not improve leave-one-year-out
	performance. The narrower one-day dummy gives the same conclusion. The dummy
	coefficients are also not systematically significant under HAC inference. Thus,
	the main benchmark is not materially affected by quarter-end dummy adjustments.
	
	This finding should not be read as evidence against the quarter-end mechanism
	in \citet{DTV18}. It reflects the maturity structure of the present sample. The
	sharp quarter-end regulatory effect is primarily a very-short-tenor phenomenon,
	whereas this paper studies government-bond CIP deviations at maturities of
	three months and longer. The benchmark is therefore best interpreted as a model
	of background government-bond CIP deviations, not ultra-short-maturity
	reporting-date spikes.
	
	\subsection{Alternative market-risk proxy: VIX}
	
	VIX is a natural alternative proxy for market-wide risk and risk-bearing
	capacity. I therefore compare the baseline with three VIX specifications. The
	first adds lagged VIX to the three baseline state variables. The second replaces
	NFCI with lagged VIX while retaining the broad dollar and Treasury slope. The
	third uses lagged VIX alone. This exercise asks whether the baseline is mainly
	capturing standard equity-market risk sentiment.
	
	\FloatBarrier
	\begin{table}[H]
		\centering
		\onehalfspacing
		\footnotesize
		\caption{Alternative VIX specifications}
		\label{tab:vix_robustness}
		
		\resizebox{\textwidth}{!}{%
			\begin{tabular}{lrrrr}
				\toprule
				Specification
				& In-sample \(R^2\)
				& LOYO pooled \(R^2\)
				& LOYO mean-year \(R^2\)
				& Expanding pooled \(R^2\) \\
				\midrule
				Baseline
				& 0.637 & 0.499 & 0.347 & 0.501 \\
				Baseline + VIX
				& 0.648 & 0.494 & 0.325 & 0.446 \\
				VIX + broad dollar + slope
				& 0.515 & 0.362 & 0.085 & 0.252 \\
				VIX only
				& 0.271 & 0.163 & -0.144 & 0.175 \\
				\bottomrule
			\end{tabular}%
		}
		
		\vspace{0.5em}
		
		\parbox{\textwidth}{%
			\footnotesize
			\textit{Notes:} The baseline uses lagged NFCI, the lagged nominal broad
			dollar index, and the lagged Treasury 10-year minus 2-year slope. VIX
			specifications use lagged VIX. LOYO denotes leave-one-year-out evaluation.
			Expanding-window forecasts use the same initial three-year training design
			as in the baseline forecast exercise.
		}
	\end{table}
	\FloatBarrier
	
	The results show that VIX contains information in isolation, but does not
	subsume the benchmark. The VIX-only specification has positive explanatory power,
	with an in-sample \(R^2\) of 0.271 and a LOYO pooled \(R^2\) of 0.163. Adding the
	broad dollar and Treasury slope raises the in-sample \(R^2\) to 0.515 and the
	LOYO pooled \(R^2\) to 0.362. These are nontrivial fits, but they remain well
	below the three-variable baseline.
	
	Adding VIX to the baseline raises the in-sample \(R^2\) only modestly, from
	0.637 to 0.648. It does not improve out-of-sample performance. The LOYO pooled
	\(R^2\) falls slightly from 0.499 to 0.494, and the expanding-window pooled
	\(R^2\) falls from 0.501 to 0.446. Thus, VIX adds some in-sample variation, but
	not stable out-of-sample information.
	
	The coefficient evidence points to the same conclusion. In the VIX-only
	specification, the VIX coefficient is positive in 91.8\% of panels and
	significant at the 5\% level in 42.9\% of panels. Once the baseline variables
	are included, however, the VIX coefficient becomes much less stable. It is
	significant in only 16.3\% of panels, with significant positive and negative
	coefficients split almost evenly. Its median one-standard-deviation effect is
	also small, about \(-0.6\) basis points. These results suggest that VIX is a
	useful standalone market-risk proxy, but its stable information for
	government-bond CIP deviations is largely absorbed by broader financial
	conditions, the dollar, and the Treasury slope.
	
	\subsection{Engle--Granger cointegration diagnostics}
	
	The high level-regression \(R^2\) raises a standard concern: persistent series
	can appear related even when there is no stable economic relation. I use
	Engle--Granger residual-based tests as a diagnostic check against this
	possibility.
	
	I run two tests. The first is a compressed actual--fitted diagnostic. It tests
	whether actual CIP deviations and baseline fitted values share a stable
	long-run relation. The second is closer to the baseline regression itself. It
	tests whether actual CIP deviations are cointegrated with the three baseline
	regressors: \(NFCI\), \(Dollar\), and \(Slope\).
	
	\Cref{tab:eg_summary} reports the rejection counts. At the currency--tenor
	level, the multivariate test using the three regressors rejects the
	no-cointegration null in 98 of 98 panels at the 5\% level under the constant
	specification. It rejects in 96 of 98 panels at the 1\% level. With a constant
	and trend, the rejection counts remain high: 97 of 98 at the 5\% level and 94
	of 98 at the 1\% level.
	
	The actual--fitted diagnostic is also strong. At the currency--tenor level, it
	rejects the no-cointegration null in all 98 panels at the 5\% level. The 1\%
	rejection count is 98 of 98 under the constant specification and 97 of 98 under
	the constant-plus-trend specification. Currency-level and global aggregate
	series reject the null in all cases, including at the 1\% level.
	
	\FloatBarrier
	\begin{table}[H]
		\centering
		\onehalfspacing
		\footnotesize
		\caption{Engle--Granger cointegration diagnostics}
		\label{tab:eg_summary}
		\begin{threeparttable}
			\setlength{\tabcolsep}{4.5pt}
			\begin{tabular}{llccc}
				\toprule
				Unit & Test relation & Deterministic term & 5\% reject & 1\% reject \\
				\midrule
				Currency--tenor panel & Actual \( \sim \) regressors & Constant & 98/98 & 96/98 \\
				Currency--tenor panel & Actual \( \sim \) regressors & Constant + trend & 97/98 & 94/98 \\
				Currency--tenor panel & Actual \( \sim \) fitted value & Constant & 98/98 & 98/98 \\
				Currency--tenor panel & Actual \( \sim \) fitted value & Constant + trend & 98/98 & 97/98 \\
				\midrule
				Currency aggregate & Actual \( \sim \) regressors & Constant & 11/11 & 11/11 \\
				Currency aggregate & Actual \( \sim \) regressors & Constant + trend & 11/11 & 11/11 \\
				Currency aggregate & Actual \( \sim \) fitted value & Constant & 11/11 & 11/11 \\
				Currency aggregate & Actual \( \sim \) fitted value & Constant + trend & 11/11 & 11/11 \\
				\midrule
				Global aggregate & Actual \( \sim \) regressors & Constant & 1/1 & 1/1 \\
				Global aggregate & Actual \( \sim \) regressors & Constant + trend & 1/1 & 1/1 \\
				Global aggregate & Actual \( \sim \) fitted value & Constant & 1/1 & 1/1 \\
				Global aggregate & Actual \( \sim \) fitted value & Constant + trend & 1/1 & 1/1 \\
				\bottomrule
			\end{tabular}
			\begin{tablenotes}
				\footnotesize
				\item Notes: Each cell reports the number of series for which the no-cointegration null is rejected.
				``Actual \( \sim \) regressors'' refers to the multivariate Engle--Granger test linking actual CIP deviations to \(NFCI\), \(Dollar\), and \(Slope\).
				``Actual \( \sim \) fitted value'' refers to the compressed Engle--Granger test between actual CIP deviations and baseline fitted values.
				Currency and global aggregates use daily median aggregate series.
			\end{tablenotes}
		\end{threeparttable}
	\end{table}
	\FloatBarrier
	
	These results reduce the concern that the baseline level fit is only a
	spurious common-trend artifact. The evidence is strongest in the multivariate
	test, where actual CIP deviations are linked directly to the three baseline
	state variables.
	
	The interpretation remains diagnostic, not causal. The Engle--Granger tests
	show stable long-run alignment. They do not show that any one regressor
	structurally causes CIP deviations. A more modest conclusion is appropriate:
	NFCI, the broad dollar, and the Treasury slope compress macrofinancial forces
	that are stably aligned with the persistent component of post-GFC
	government-bond CIP deviations.
	
	\subsection{Non-overlapping aggregation-difference analysis}
	
	The next diagnostic asks whether the benchmark only explains persistent levels.
	For each currency--tenor panel, I sort observations by date and form
	non-overlapping \(N\)-trading-day blocks. I average the CIP deviation and the
	three regressors within each block, then difference adjacent blocks:
	\[
	\Delta_N CIP_{c,\tau,b}
	=
	a_{c,\tau}
	+
	b^N_{c,\tau}\Delta_N NFCI_b
	+
	b^D_{c,\tau}\Delta_N Dollar_b
	+
	b^S_{c,\tau}\Delta_N Slope_b
	+
	e_{c,\tau,b}.
	\]
	Here \(b\) indexes non-overlapping aggregation blocks. This design tests whether
	the relation appears in changes measured over horizons from one day to roughly
	three months. The non-overlapping construction reduces mechanical serial
	correlation from overlapping differences, but the effective sample size falls
	as \(N\) increases.
	
	\Cref{tab:diff_summary} reports the results for selected aggregation windows.
	The pattern is simple. At high frequencies, the benchmark explains almost
	nothing. At intermediate horizons, the relation reappears. At longer horizons,
	the pooled evidence remains positive, but annual panel-level evaluation becomes
	noisy because the number of non-overlapping blocks is small.
	
	For \(N=1\) and \(N=5\), explanatory power is close to zero. The in-sample
	\(R^2\) is 0.007 at \(N=1\) and 0.047 at \(N=5\). The LOYO pooled \(R^2\) is
	also near zero. This is consistent with the design of the benchmark. Three
	lagged macrofinancial state variables are not meant to predict daily execution
	noise, flow imbalances, or dealer inventory shocks.
	
	The relation becomes stronger at monthly-to-quarterly horizons. The in-sample
	\(R^2\) rises to 0.218 at \(N=21\), 0.313 at \(N=32\), and 0.369 at \(N=50\).
	The LOYO pooled \(R^2\) is also positive over these windows, reaching 0.254 at
	\(N=50\). Thus, the benchmark explains not only persistent levels, but also a
	meaningful share of lower-frequency changes.
	
	The LOYO mean-year \(R^2\) becomes negative at longer \(N\). This statistic
	should be interpreted cautiously. The number of differenced observations falls
	from 419,715 at \(N=1\) to 6,505 at \(N=63\). Once these observations are split
	across 98 panels and calendar-year holdouts, each year-by-panel cell contains
	only a small number of non-overlapping blocks. The mean-year statistic is
	therefore high variance. It can turn negative even when the pooled \(R^2\)
	remains positive.
	
	\FloatBarrier
	\begin{table}[H]
		\centering
		\onehalfspacing
		\footnotesize
		\caption{Non-overlapping aggregation-difference performance}
		\label{tab:diff_summary}
		\begin{threeparttable}
			\setlength{\tabcolsep}{5pt}
			\begin{tabular}{rrrrr}
				\toprule
				\makecell{Aggregation\\window \(N\)} 
				& \makecell{In-sample\\\(R^2\)} 
				& \makecell{LOYO pooled\\\(R^2\)} 
				& \makecell{LOYO mean-year\\\(R^2\)} 
				& Observations \\
				\midrule
				1  & 0.007 & -0.001 &  0.004 & 419,715 \\
				5  & 0.047 &  0.026 & -0.006 &  83,823 \\
				21 & 0.218 &  0.161 & -0.045 &  19,848 \\
				32 & 0.313 &  0.234 & -0.049 &  12,951 \\
				50 & 0.369 &  0.254 & -0.060 &   8,232 \\
				63 & 0.412 &  0.217 & -0.115 &   6,505 \\
				\bottomrule
			\end{tabular}
			\begin{tablenotes}
				\footnotesize
				\item Notes: For each \(N\), I form non-overlapping \(N\)-trading-day block averages and difference adjacent blocks.
				In-sample \(R^2\) uses a within-panel denominator based on panel-specific differenced means.
				LOYO \(R^2\) uses the training-sample panel-specific differenced mean as the benchmark.
				Observations report the total number of differenced observations across all currency--tenor panels.
			\end{tablenotes}
		\end{threeparttable}
	\end{table}
	\FloatBarrier
	
\begin{figure}[H]
		\centering
		\includegraphics[width=6.5in]{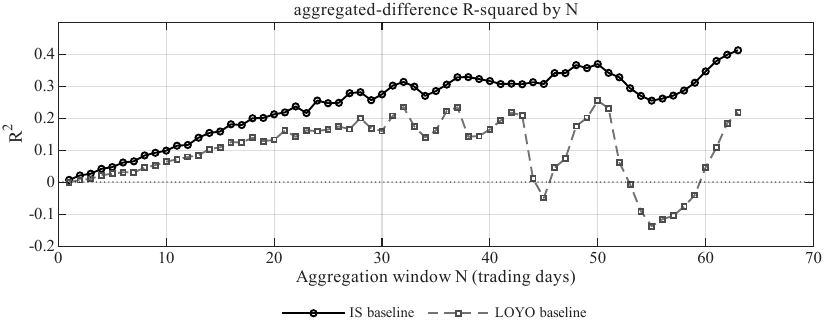}
		\caption{Non-overlapping aggregation-difference performance}
		\label{fig:diff}
	\end{figure}

	\Cref{fig:diff} plots the in-sample and out-of-sample \(R^2\) values across
	aggregation windows. The figure reinforces the table. The benchmark explains
	little of daily changes. Its explanatory power rises once the horizon reaches
	several weeks. The pooled LOYO curve tracks the in-sample curve at intermediate
	windows and remains positive throughout.
	
	\FloatBarrier
	\FloatBarrier
	
	The aggregation-difference evidence sharpens the interpretation of the
	benchmark. The relation is weak at daily horizons, meaningful at
	monthly-to-quarterly horizons, and harder to estimate precisely at long
	non-overlapping windows. This is consistent with a persistent macrofinancial
	component rather than a high-frequency arbitrage-execution model.
	
	\subsection{Principal component analysis}
	\label{subsec:pca}
	
	As a final diagnostic, I rotate the three baseline state variables into
	principal components. This asks whether the benchmark is mostly a proxy for one
	or two broad common factors, or whether the full three-variable state space
	matters. The PCA is applied to the standardized baseline regressors: lagged
	NFCI, the lagged nominal broad dollar index, and the lagged Treasury 10-year
	minus 2-year slope.
	
	\Cref{tab:pca_robustness_fit} reports the fit. The first principal component
	explains 64.1\% of the variation in the state variables, but it delivers only
	an in-sample \(R^2\) of 0.353 and a negative expanding-window pooled \(R^2\).
	Adding the second component raises the in-sample \(R^2\) to 0.613, but the
	expanding-window pooled \(R^2\) remains low at 0.033. The full
	three-component rotation reproduces the baseline by construction and restores
	the expanding-window pooled \(R^2\) of 0.503.
	
	\FloatBarrier
	\begin{table}[H]
		\centering
		\onehalfspacing
		\footnotesize
		\caption{PCA robustness: fit}
		\label{tab:pca_robustness_fit}
		\begin{threeparttable}
			\begin{tabular}{lrrrr}
				\toprule
				Model & Cumulative state variance & In-sample \(R^2\) &
				Expanding pooled \(R^2\) & Mean yearly pooled \(R^2\) \\
				\midrule
				PC1 & 64.1\% & 0.353 & -0.043 & 0.040 \\
				PC1--PC2 & 95.1\% & 0.613 & 0.033 & 0.112 \\
				PC1--PC3 & 100.0\% & 0.637 & 0.503 & 0.495 \\
				Baseline & -- & 0.637 & 0.503 & 0.495 \\
				\bottomrule
			\end{tabular}
			\begin{tablenotes}
				\footnotesize
				\item Notes: The PCA is computed from standardized baseline state variables. The first three rows report fit using the first \(k\) principal components as regressors. The full three-component rotation spans the same regressor space as the baseline, so its fit matches the baseline up to numerical precision.
			\end{tablenotes}
		\end{threeparttable}
	\end{table}
	\FloatBarrier
	
	\Cref{tab:pca_loadings} reports the loadings. PC1 mainly contrasts the broad
	dollar with the Treasury slope. PC2 is dominated by NFCI. PC3 loads positively
	on both the broad dollar and the Treasury slope. Although PC3 explains only a
	small share of state-variable variance, it is needed to recover the benchmark's
	out-of-sample performance.
	
	\FloatBarrier
	\begin{table}[H]
		\centering
		\onehalfspacing
		\footnotesize
		\caption{PCA loadings}
		\label{tab:pca_loadings}
		\begin{threeparttable}
			\begin{tabular}{lrrr}
				\toprule
				State variable & PC1 & PC2 & PC3 \\
				\midrule
				NFCI & -0.263 & 0.965 & 0.018 \\
				Broad dollar & 0.684 & 0.174 & 0.709 \\
				Treasury 10Y--2Y slope & -0.681 & -0.199 & 0.705 \\
				\bottomrule
			\end{tabular}
			\begin{tablenotes}
				\footnotesize
				\item Notes: Loadings are computed from the standardized baseline state variables. The sign of each principal component is arbitrary.
			\end{tablenotes}
		\end{threeparttable}
	\end{table}
	\FloatBarrier
	
	The PCA evidence rules out a simple common-factor interpretation. The first two
	components explain 95.1\% of state-variable variation, but they do not recover
	the expanding-window performance of the original benchmark. The low-variance
	third component is needed. The benchmark therefore does not appear to be a
	redundant proxy for one or two broad factors. Its performance depends on the
	full combination of NFCI, the broad dollar, and the Treasury slope.
	
	Taken together, the robustness tests support the benchmark interpretation. The
	results are not explained by a short-tenor quarter-end dummy. VIX contains
	standalone information, but it does not subsume the three-variable benchmark and
	does not improve out-of-sample performance once the baseline state variables are
	included. The level relation passes residual-based cointegration diagnostics.
	The relation appears in lower-frequency changes, but not in daily noise. The PCA
	results show that the full three-variable state space matters. The benchmark
	should therefore be read as a compact public-data model of the persistent
	macrofinancial component of post-GFC government-bond CIP deviations.
	
	% =================================================
	% Section: Discussion
	% =================================================
	\section{Discussion}
	\label{sec:discuss}
	
	The results should be read as a benchmark result, not as a structural model of
	CIP deviations. The baseline uses three public state variables. NFCI summarizes
	broad financial conditions. The broad dollar index summarizes the global dollar
	regime. The Treasury 10-year minus 2-year slope summarizes the U.S. yield-curve
	environment. These variables compress several macrofinancial forces, but they do
	not identify a single mechanism. Nor do they imply that only three forces drive
	government-bond CIP deviations.
	
	The central result is narrower. A small public-data specification explains a
	large persistent component of daily government-bond CIP deviations. The model is
	easy to replicate and strong enough that it should not be treated as a straw
	man. This section discusses how the benchmark should be used, where it performs
	less well, and how it relates to richer explanations.
	
	\subsection{How the benchmark should be used}
	
	The benchmark is meant to be a hurdle for future empirical work. A researcher
	who proposes a new funding variable, regulatory proxy, liquidity measure,
	balance-sheet variable, hedging-demand proxy, or structural model should ask
	whether it improves on this baseline. Statistical significance in isolation is
	not enough. The relevant test is incremental explanatory power relative to a
	compact public-data specification.
	
	This comparison can be implemented as a horse race. A richer model should be
	tested against the baseline both in sample and out of sample. It should also
	show that the improvement is not driven by one currency, one tenor, or one
	favorable holdout year. In this sense, the baseline provides a common yardstick
	for daily government-bond CIP regressions.
	
	This role is useful because many important CIP determinants are hard to observe
	at daily frequency. Dealer balance-sheet capacity, bank-level constraints,
	specialized liquidity measures, regulatory exposures, and hedging flows are
	often proprietary, low-frequency, or difficult to reconstruct. The public
	benchmark does not replace these variables. It gives them a transparent
	comparison model.
	
	\subsection{Currency-level heterogeneity}
	
	The baseline is deliberately dollar-centered. This is natural for CIP
	deviations, but it also limits what the model can explain. NFCI, the broad
	dollar, and the Treasury slope are U.S. or global-dollar state variables. Their
	effects need not be identical across currencies. Currency-specific funding
	structures, hedging demand, local bond markets, safe-asset roles, and domestic
	institutions can all affect the mapping from global state variables to CIP
	deviations.
	
	The results show this heterogeneity. CAD, EUR, KRW, DKK, and SEK perform well in
	the currency-level common-slope specification. JPY is weaker. The model captures
	part of the common movement in JPY, but it misses several level shifts,
	especially in later holdout years. This is consistent with a larger local
	component that is not spanned by the three U.S.-centered state variables.
	
	The weaker cases are informative rather than fatal. Some currencies are
	difficult only at specific maturities. For example, performance is weaker for
	some long-tenor SEK panels and some short- or intermediate-tenor AUD panels.
	This suggests that the missing component is partly currency-specific and partly
	tenor-specific. Local curve conditions, maturity-specific hedging pressure, and
	segmented investor demand may matter in ways that the baseline does not capture.
	
	This does not undermine the benchmark interpretation. A useful benchmark should
	be portable and disciplined. It should not absorb every local market feature.
	Weak currency--tenor cells show where richer models have room to add value. A
	successful extension should explain these residual patterns, not only raise
	pooled \(R^2\).
	
	\subsection{Regime dependence and frequency scope}
	
	The out-of-sample results also show regime dependence. The benchmark performs
	well in most holdout years, but not equally in every year. This is expected.
	Government-bond CIP deviations reflect balance-sheet capacity, dollar funding,
	hedging imbalances, regulation, safe-asset demand, and market liquidity. The
	relative importance of these forces can change across regimes.
	
	The quarter-end dummy exercise clarifies the maturity scope of the paper.
	\citet{DTV18} emphasize sharp quarter-end regulatory effects, especially at very
	short maturities. The data used here begin at the three-month tenor. A
	short-tenor quarter-end dummy adds essentially no explanatory power to the
	baseline. This does not contradict the quarter-end mechanism. It means that the
	present benchmark is aimed at the persistent background component of
	government-bond CIP deviations, not at ultra-short-maturity reporting-date
	spikes.
	
	The aggregation-difference results point to the same interpretation. The
	baseline explains little of daily changes. Its explanatory power rises when
	changes are measured over several weeks or months. The model should therefore
	not be interpreted as a high-frequency execution model. It is better understood
	as a compact description of the low-frequency macrofinancial component of
	post-GFC government-bond CIP deviations.
	
	\subsection{Relation to richer macrofinancial explanations}
	
	The benchmark complements the macrofinancial determinants literature. It does
	not replace it. \citet{COZ21}, for example, relate cross-currency bases to FX
	market liquidity, intermediary risk-taking capacity, broad dollar strength,
	central bank balance sheets, and term premia. Other work emphasizes bank
	balance sheets, regulatory constraints, dollar funding stress, quarter-end
	effects, and hedging imbalances. These mechanisms remain central for
	understanding why CIP deviations exist.
	
	The contribution here is to ask how far one can go before using lower-access
	variables. The answer is: surprisingly far, but not all the way. The three
	public state variables capture a large common component. They leave meaningful
	residual heterogeneity across currencies, tenors, and regimes. That residual
	heterogeneity is where structural work and richer data should matter.
	
	Overall, post-GFC government-bond CIP deviations are not purely idiosyncratic
	pricing errors. They are persistent price variables aligned with public
	macrofinancial conditions. The alignment is strong but incomplete. The
	combination of NFCI, the broad dollar index, and the Treasury slope therefore
	provides a simple, reproducible, and empirically strong benchmark for future
	work on daily government-bond CIP deviations.
	
	% =================================================
	% Section: Conclusion
	% =================================================
	\section{Conclusion}
	\label{sec:conclusion}
	
	This paper proposes a compact public-data benchmark for post-GFC
	government-bond covered interest parity (CIP) deviations. The motivation is
	practical. CIP deviations are observed at daily frequency, but the empirical
	CIP literature has not converged on a canonical daily-frequency benchmark.
	Many important determinants---intermediary balance-sheet constraints, dollar
	funding conditions, regulation, hedging demand, liquidity, and term premia---are
	lower-frequency, difficult to reconstruct, or not equally portable across
	research environments. The paper asks how far one can go before using these
	richer variables.
	
	The answer is: surprisingly far. A baseline with only three lagged public state
	variables---the Chicago Fed NFCI, the nominal broad U.S. dollar index, and the
	Treasury 10-year minus 2-year yield slope---summarizes a large component of
	G10+KRW government-bond CIP deviations. The fit is strong in separate
	currency--tenor regressions, in currency-level common-slope specifications, and
	in pooled common-slope specifications. It also remains strong in
	leave-one-year-out out-of-sample tests. The result is therefore not merely an
	in-sample artifact of a flexible panel specification.
	
	The coefficient patterns are consistent with a macrofinancial interpretation.
	NFCI is the most stable regressor and enters positively in most panels,
	consistent with tighter financial conditions widening CIP deviations. The
	Treasury slope also tends to enter positively, suggesting that the U.S.
	yield-curve regime is linked to the persistent component of government-bond CIP
	deviations. The broad dollar coefficient is more heterogeneous, consistent with
	the global dollar cycle interacting with currency-specific funding and hedging
	pressures.
	
	The robustness exercises clarify the scope of the benchmark. A short-tenor
	quarter-end dummy adds essentially no explanatory power. This does not
	contradict the quarter-end mechanism emphasized by \citet{DTV18}; it reflects
	the maturity structure of the present data, which begin at the three-month
	tenor. Engle--Granger diagnostics reduce the concern that the level fit is only
	a common-trend artifact. Aggregation-difference tests show that the benchmark
	explains little of daily high-frequency changes, but becomes more informative
	over several-week to several-month horizons. The specification is therefore not
	a high-frequency arbitrage-execution model. It is a reduced-form benchmark for
	the persistent macrofinancial component of government-bond CIP deviations.
	
	The contribution is not a new individual state variable or a structural
	identification result. NFCI, the broad dollar, and the Treasury slope are all
	motivated by existing work. The contribution is to show that their combination
	forms a simple, reproducible, and empirically strong daily benchmark. This
	benchmark provides a common hurdle for future work. Richer funding variables,
	regulatory proxies, intermediary balance-sheet measures, liquidity measures,
	hedging-demand variables, or structural models should show incremental power
	relative to it, both in sample and out of sample, and across currencies and
	tenors.
	
	In sum, post-GFC government-bond CIP deviations are not purely idiosyncratic
	pricing errors. They are persistent price variables aligned with observable
	macrofinancial conditions. This paper does not replace structural models of CIP
	deviations. It provides a portable public-data benchmark for the background
	component that such models should be compared against.
	
	\subsection*{Funding}
	This research did not receive any specific grant from funding agencies in the public, commercial, or not-for-profit sectors.
	
	\subsection*{Declaration of AI usage in manuscript preparation}
	During the preparation of this manuscript, the author used ChatGPT (OpenAI) and Claude (Anthropic) for language refinement and structural clarity.
	All outputs were reviewed and edited by the author, who takes full responsibility for the content.
	
	\subsection*{Declaration of interest}
	The author declares no competing interests.
	
	% =================================================
	\newpage
	\onehalfspacing

\end{document}